\documentclass[conference]{IEEEtran}

\usepackage{amsmath,graphicx}
\usepackage{times}
\usepackage{epsfig}
\usepackage{url} 
\usepackage{booktabs} 
\usepackage{amsfonts} 
\usepackage{amssymb} 
\usepackage{nicefrac} 
\usepackage{microtype}
\usepackage{mathrsfs}
\usepackage{bm} 
\usepackage{cite} 
\usepackage{comment}
\usepackage{graphicx} 
\usepackage{mathtools}
\usepackage{algpseudocode}
\usepackage[ruled,linesnumbered]{algorithm2e}
\usepackage{amsthm} 
\usepackage{enumitem} 
\usepackage{multirow} 
\usepackage{tabularx} 
\usepackage{hhline}
\usepackage{arydshln} 
\usepackage{enumitem} 
\usepackage{siunitx}
\usepackage[font=small,skip=5pt]{caption}
\usepackage{comment}
\usepackage[dvipsnames]{xcolor}
\usepackage{hyperref}

\newcolumntype{L}[1]{>{\raggedright\arraybackslash}p{#1}}
\newcolumntype{C}[1]{>{\centering\arraybackslash}p{#1}}
\newcolumntype{R}[1]{>{\raggedleft\arraybackslash}p{#1}}

\theoremstyle{plain} 

\newtheorem{assumption}{Assumption}

\def\argmin{\mathop{\mathsf{arg\,min}}} 

\def\lim{\mathop{\mathsf{lim}}} 

\def\prox{\mathsf{prox}}

\def\ebm{{\bm{e}}}

\def\sbm{{\bm{s}}}
\def\xbm{{\bm{x}}}

\def\ybm{{\bm{y}}}
\def\zbm{{\bm{z}}}

\def\mbm{{\bm{m}}}

\def\thetabm{{\bm{\theta}}}

\def\Hbm{{\bm{H}}}

\def\Fbm{{\bm{F}}}
\def\Sbm{{\bm{S}}}
\def\Pbm{{\bm{P}}}

\def\xbmhat{{\widehat{\bm{x}}}}

\def\xbfhat{{\widehat{\mathbf{x}}}}
\def\xbfhat{{\widehat{\bm{x}}}}

\def\C{\mathbb{C}}
\def\R{\mathbb{R}}

\def\Rcal{{\mathcal{R}}}

\def\Gcal{{\mathcal{G}}}

\def\Lcal{{\mathcal{L}}}

\begin{document}

\title{Image Reconstruction for MRI using Deep CNN Priors Trained without Groundtruth}

\author{\IEEEauthorblockN{
Weijie Gan\IEEEauthorrefmark{1},
Cihat Eldeniz\IEEEauthorrefmark{2},
Jiaming Liu\IEEEauthorrefmark{3}, 
Sihao Chen\IEEEauthorrefmark{4},
Hongyu An\IEEEauthorrefmark{2} and
Ulugbek S.~Kamilov\IEEEauthorrefmark{1}\IEEEauthorrefmark{3}}
\vspace{4pt}

\IEEEauthorblockA{\IEEEauthorrefmark{1} Department of Computer Science and Engineering,
Washington University in St. Louis, MO, USA}
\IEEEauthorblockA{\IEEEauthorrefmark{2} Mallinckrodt Institute of Radiology,
Washington University in St. Louis, MO, USA}
\IEEEauthorblockA{\IEEEauthorrefmark{3} Department of Electrical and Systems Engineering, 
Washington University in St. Louis, MO, USA}
\IEEEauthorblockA{\IEEEauthorrefmark{4} Department of Biomedical Engineering,
Washington University in St. Louis, MO, USA}
\small{Emails}: \texttt{\{weijie.gan, cihat.eldeniz, jiaming.liu, sihaochen, hongyuan, kamilov\}@wustl.edu}
}

\maketitle

\begin{abstract}
We propose a new plug-and-play priors (PnP) based MR image reconstruction method that systematically enforces data consistency while also exploiting deep-learning priors. Our prior is specified through a convolutional neural network (CNN) trained without any artifact-free ground truth to remove undersampling artifacts from MR images. The results on reconstructing free-breathing MRI data into ten respiratory phases show that the method can form high-quality 4D images from severely undersampled measurements corresponding to acquisitions of about 1 and 2 minutes in length. The results also highlight the competitive performance of the method compared to several popular alternatives, including the TGV regularization and traditional UNet3D.
\end{abstract}

\begin{IEEEkeywords}
Image reconstruction, magnetic resonance imaging, plug-and-play priors, deep learning.
\end{IEEEkeywords}

\section{Introduction}
Magnetic resonance imaging (MRI) is a widely used non-invasive imaging technology.
However, MRI typically requires long scanning time for obtaining high quality images.
Recently, there has been interest in reducing MRI acquisition times.
One approach is \emph{compressive sensing magnetic resonance imaging (CS-MRI)}~\cite{lustig2007sparse}, which aims to reconstruct unknown images $\xbm$ from sparsely-sampled Fourier measurements $\ybm$. The reconstruction is often formulated as an optimization problem
\begin{equation}
  \label{equ:prob}
  \xbfhat = \argmin_{\xbm} \big\{g(\xbm) + h(\xbm)\big\}\ ,
\end{equation}
where $g$ is the data-fidelity term that penalizes the mismatch to the measurements and $h$ is the regularizer that imposes prior knowledge regarding the unknown image. Over the years, many regularizers have been proposed in the context of image reconstruction, including those based on transform-domain sparsity, low-rank penalty, and dictionary learning~\cite{knoll2011second, figueiredo2001wavelet, degraux2017online}. Additionally, a variety of proximal algorithm~\cite{parikh2014proximal} have been developed to handle large amount of data and nondifferentiable regularizers. Recently, Venkatakrishnan \emph{et al.}~\cite{venkatakrishnan2013plug} introduced a powerful \emph{plug-and-play priors (PnP)} framework that leverages the mathematical equivalence between proximal operators and image denoisers to exploit powerful denoisers as imaging priors.
Common choices of denoisers include BM3D~\cite{dabov2007image}, TNRD~\cite{Chen.Pock2016} and several \emph{convolutional neural network (CNN)} architectures~\cite{Ryu.etal2019, Ahmad.etal2019, Xu.etal2020, Song2020}. Similar idea has also been used in a related class of algorithms known as \emph{regularization by denoising (RED)}~\cite{Romano.etal2017,Metzler.etal2018,Sun.etal2019c}.
It has been shown that when equipped with advanced CNN priors, these methods provide excellent performance because of their ability to exploit information from both the deep learning model and physics-based forward model. However, the training of corresponding priors can be a significantly practical challenge for CS-MRI, where the measurement setup fundamentally limits the ability to collect fully-sampled groundtruth data. This limitation has also motivated research on unsupervised DL schemes that rely exclusively on information available in corrupted measurements\cite{Lehtinen2018, Yaman2020, Gan2020}. One widely used technique is \emph{Noise2Noise (N2N)}~\cite{Lehtinen2018} that trains a CNN by mapping pairs of observations of the same image containing different noise realizations. However, it is still challenging to obtain multiple accurately calibrated measurements from the same subject~\cite{Gan2020}.

In this paper, we proposed a new PnP based MRI reconstruction method called \emph{Multi-Scale learning for MRI-Plug and Play (MSMRI-PnP)}. MSMRI-PnP can systematically enforce data consistency of the MRI system while exploiting a CNN prior.
The key contributions of this paper are two-fold: 
(1) Our method exploits CNN priors trained for removing streaking artifacts as well as noise from an undersampled MRI data. Inspired by N2N, the CNN is uses downsampled variants of a single artifact corrupted image, reducing the dependency on fully-sampled data or multiple calibrated measurements from the same subject. This CNN is called the \emph{Multi-Scale learning for MRI Network (MSMRI-Net)}. (2) We validate MSMRI-PnP on in-vivo acquired data by reconstructing 4D MR images from highly undersampled k-space measurements and highlight promising results compared against purely model-based or CNN-based approaches.

\begin{figure*}[!t]
  \label{fig:method} 
  \centering
  \includegraphics[width=18cm]{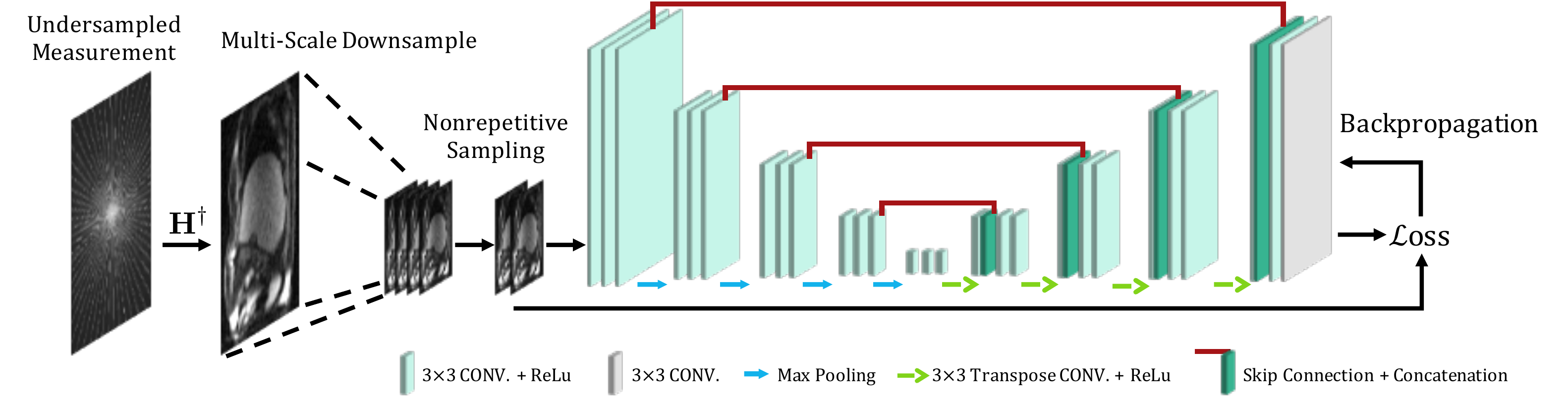}
  \caption{
    Illustration of the proposed MSMRI-Net that trains a CNN on a single undersampled measurements without corresponding fully-sampled groundtruth.
    We apply the pseudoinverse of the MRI forward operator on the measurements to obtain a corrupted backprojection in the image domain. 
    One can then generate downsampled variants of the corrupted image by applying the \emph{multi-scale downsampling} (see text for more details) operator.
    The CNN is trained by mapping the downsampled variants to each other.
    The details of CNN architecture are also provided.
  }
  \label{fig:method}
\end{figure*}
\section{Related Work}
\subsection{Inverse Problem Formulation}
Consider the linear inverse problem of recovering an unknown image $\xbm\in\C^n$ from its undersampled k-space measurements $\ybm\in\C^m$
\begin{equation}
  \ybm = \Hbm\xbm + \ebm\ ,
\end{equation}
where $\ebm\in\C^m$ is a noise vector and $\Hbm\in\C^{m\times n}$ is a measurement operator. 
For example, in dynamic parallel CS-MRI~\cite{Knoll2019,Deshmane2012a} with radial sampling, the forward model $\Hbm$ can be formulated as
\begin{equation}
  \Hbm_i^{(t)} = \Pbm^{(t)}\Fbm\Sbm_i\ ,
\end{equation}
where $\Fbm$ is the nonuniform fast Fourier transform (NUFFT) operator, $\Sbm_i$ denotes the matrix of the element-wise sensitivity map of the $i$th coil, and $\Pbm^{(t)}$ represents the k-space sampling operator at time $t$. One can then represents the measurements $\ybm$ as a 4D matrix $\ybm(k, m, n, t)$ with $t$ as respiratory phases and $(k, m, n)$ refers to three spacial directions. The inverse problem of the CS-MRI is often ill posed, requiring the use of the formulation such as in~\eqref{equ:prob}. In such cases, one of the most popular data-fidelity terms is least-squares 
\begin{equation}
\label{equ:l2}
g(\xbm) = \frac{1}{2}\|\ybm - \Hbm\xbm\|^2_2,
\end{equation} 
which imposes an $\ell_2$-penalty on the data-fit. Many popular regularizers, such as the ones based on the $\ell_1$-norm, are non-differentiable. A variety of methods such as the proximal gradient method (PGM) and the alternating direction method of multipliers (ADMM)~\cite{Boyd.etal2011} have been developed for efficient minimization of nonsmooth functions, without differentiating them, by using the~\emph{proximal operator} defined as
\begin{equation}
\label{equ:prox}
\prox_{\mu h}(\zbm) \triangleq \argmin_{\xbm\in\R^N}\Big\{\frac{1}{2}||\xbm-\zbm||_2^2+\mu h(\xbm)\Big\}\ ,
\end{equation} 
where $\mu > 0$ is a parameter. The definition~\eqref{equ:prox} implies that the proximal operator corresponds to an image denoiser formulated as regularized optimization. Additionally, since the proximal operator is mathematically equivalent to regularized image denoising,  the powerful idea of the PnP~\cite{venkatakrishnan2013plug} was to consider replacing it with an arbitrary image denoiser $D_{\sigma}$, where $\sigma > 0$ controls the strength of denoising. One can then solve it by running the following fixed-point algorithm
\begin{equation}
  \label{equ:pnp}
  \xbm^k \leftarrow D_{\sigma}(\xbm^{k-1}-\gamma\nabla g(\xbm^{k-1})),\
\end{equation} 
where $\gamma > 0$ is the step-size, and $\nabla g$ denotes the gradient of the data-fit. The key advantage of the PnP is that it can regularize the CS-MRI reconstruction by using different off-the-shelf denoisers.

\subsection{Deep Learning Algorithms for CS-MRI}
Recently, deep learning has gained popularity for solving the CS-MRI problem. 
The traditional deep learning methods rely on fully-sampled data as training targets for CNNs.
Despite that, such groundtruth images are not always available in practice, which limits the usage of these methods.
N2N~\cite{Lehtinen2018} is a recent DL methodology to address this challenges that trains a CNN without need of groundtruth.
It considers a group of calibrated noisy images $\{\xbmhat_{ij}\}$ where $j$ indexes denote different realizations of the same underlying image $i$.
In the context of the CS-MRI, $ij$ might refer the $j^{\text{th}}$ MRI acquisition of the Subject $i$, with each acquisition consisting of an independent sampling pattern and measurement noises. 
The CNN in the N2N can be trained via the following minimization
\begin{equation}
  \argmin_\thetabm \sum_{i, j,j'}\ \Lcal\big(\mathsf{f}_\thetabm(\xbmhat_{ij}),\ \xbmhat_{ij'}\big)\ ,
\end{equation}
where $\mathsf{f}_\thetabm$ is the CNN parametrized by $\thetabm$ and $\Lcal$ is the loss function. 
While N2N can reduce dependency of DL on fully-sampled measurements, it complicates training of CNN due to its requirement of multiple and calibrated MRI acquisitions of a same subject.

\begin{figure*}[!t]
  \centering
  \includegraphics[width=18cm]{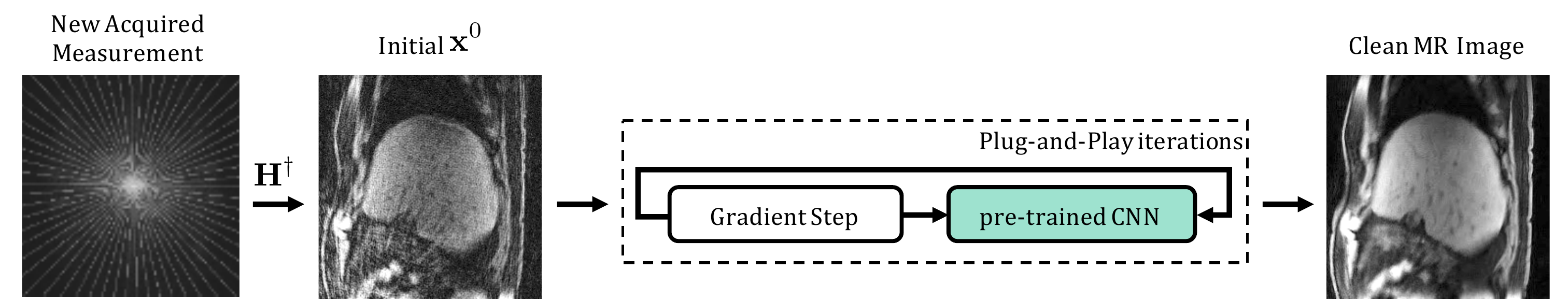}
  \caption{
    Illustration of the proposed MSMRI-PnP framework on the problem of 4D MRI reconstruction. 
    The MSMRI-PnP takes a zero-filled image as an initial vector and provides a clean reconstruction.
    This framework is capable of combining information from a pre-trained MSMRI-Net prior with that from the physics of MRI measurement operator to iteratively refine the solution within the PnP algorithm.
  }
  \label{fig:inference}
\end{figure*}

\section{Method}
\subsection{Multi-Scale Learning for MRI Network (MSMRI-Net)}
In this section, we introduce the proposed MSMRI-Net, which exclusively relies on a single undersampled MRI acquisition. 
\begin{assumption}\label{assum}
 The underlying assumption in MSMRI-Net is that the noise in neighboring pixels is independent.
\end{assumption}

Specifically, let $\xbmhat(m,n)$ denotes a pixels of an image $\xbmhat$ with $m$ and $n$ index width and height dimensions, respectively. 
The assumption indicates that measurement noise values are uncorrelated in any set of 2 by 2 regions $\Rcal_{\xbmhat}(m,n)=\big\{\xbmhat(m,n), \xbmhat(m+1,n), \xbmhat(m,n+1), \xbmhat(m+1,n+1)\big\}$. 
Here, we define $m=1,3,5,...,\ M-1$ with $M$ being the width of $\xbmhat$ and $n=1,3,5,...,\ N-1$ with $N$ being the height of $\xbmhat$. 
One can then reorganize elements with same index of each $\Rcal_{\xbmhat}$ into a smaller-scale image with its width as $M / 2$ and height as $N / 2$ corresponding to $\xbmhat$.
In particular, if we manipulated all the \emph{first} elements of $\Rcal_{\xbmhat}$ to obtain a smaller variant of $\xbfhat$, it would be equivalent to applying a downsampling operation on $\xbmhat$ by using $odd$ indexes of both the width and height dimensions.
We then termed this smaller image as $\xbmhat_\text{odd,odd}$.
Furthermore, for any $\xbmhat$, we can define a set containing four its downsampled variants as $\Gcal_{\xbmhat}=\{\xbmhat_\text{odd,odd}, \xbmhat_\text{odd,even}, \xbmhat_\text{even,odd}, \xbmhat_\text{even,even}\}$.
Here, though we take assumption based on 2 by 2 region, one can easily generalize this assumption on $n$ by $n$ regions with $n > 1$ and obtain a set of downsampled variants of $\xbmhat$ with their width as $M / n$ and height as $N / n$. 
Due to the feasibility to generate different scale downsampled variant, we termed our way to obtain $\Gcal_{\xbmhat}$ as \emph{multi-scale downsampling}.

The MSMRI-Net pipeline is summarized as Fig.\  \ref{fig:method}. 
With a undersampled measurement $\ybm$ as an exclusive training sample, a CNN is trained by mapping downsampled variants $\Gcal_{\xbmhat}$ of the corresponding zero-filled image to each other.
Here, the key idea is to manipulate the fact that measurement noise values of images in $\Gcal_{\xbmhat}$ are elementary-wise uncorrelated with each other. 
Thus, those four images can be considered as different noisy realizations of a same unknown downsampled image $\xbmhat$.
Our training procedure is conducted on a dataset containing multiple undersampled measurements from different subjects and by minimizing the following empirical risk:
\begin{equation}
  \argmin_\thetabm \sum_i\sum_{\mbm,\mbm'}^{\Gcal_{\xbmhat}^i}\ \Lcal(\mathsf{f}_\thetabm(\mbm), \mbm')
\end{equation}
where $i$ indexes different training sample and $\mathsf{f}_\thetabm$ is the CNN parametrized by $\thetabm$. Here, $\mbm$ and $\mbm'$ are two different sample of a \emph{multi-scale downsampling} set $\Gcal_{\xbmhat}^i$ corresponding to the $i$th zero-filled image $\xbmhat^i$. The CNN $\mathsf{f}_\thetabm$ is customized from widely used UNet architecture~\cite{Ronneberger2015}.

Since we consider 4D MRI reconstruction in this work, we have experiments with different ways of implementing 3D MSMRI-Net on 4D data.
Each 4D image $\xbmhat(k,m,n,t)$ with $t$ as the respiratory phase and $(k,m,n)$ as spatial dimensions would be decomposed into a succession of 3D volumes $\xbmhat(m,n,t)$. 
We then apply \emph{multi-scale downsampling} exclusively on $(m, n)$ dimension while keep $t$ dimension completed.
The CNN takes 3D downsampled variants as input and our reconstruction for whole 4D data is done slice-by-slice.

\begin{algorithm}
  \SetKwInOut{Input}{input}
  \Input{$\xbm^0=\sbm^0\in \mathbb{R}^n$ and $\gamma>0$, $\{q_k\}_{k\in\mathbb{N}}$}

  \For{${\bm k}=1,2,...$}{
      $\zbm^0 \leftarrow \sbm^{k-1} - \gamma \nabla g(\sbm^{k-1})$ \\
      $\ \ \ \ \text{where } \nabla g(\xbm) = \Hbm^\dagger(\Hbm \xbm - \ybm)$\\
      $\xbm^k \leftarrow \textit{MSMRI-Net}(\zbm^k)$ \\
      $q_k \leftarrow \frac{1}{2}(1 + \sqrt{1+4q^2_{k-1}})$ \\
      $\sbm^k \leftarrow \xbm^k + ((q_{k-1} - 1) / q_k))(\xbm^k-\xbm^{k-1})$ \\
      }
      
  \caption{MSMRI-PnP}
  \label{algo}

\end{algorithm}

\begin{figure*}[!t]
  \centering
  \includegraphics[width=18cm]{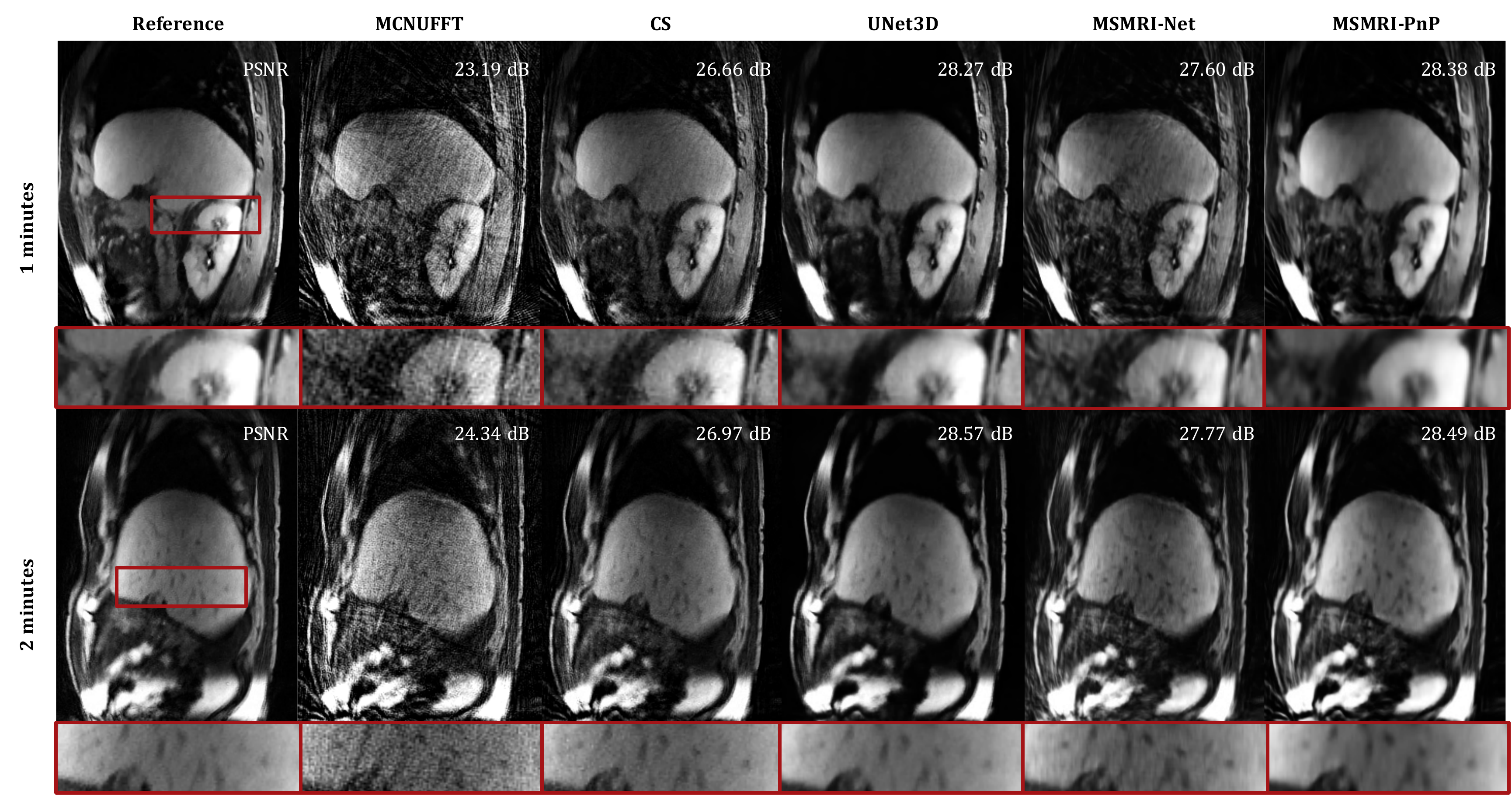}
  \caption{Illustration of several reconstruction methods on undersampled 4D MRI data of 1 minute (top) and 2 minute (bottom) acquisitions. The top-right corner of each image provides the PSNR values in reference to the reference. CS denotes model-based algorithm by using TGV as an image prior. Reference indicates the CS reconstruction of the 5 minute acquisition, which serves as a clean reference in this cases. While UNet3D depends on Reference reconstructed from measurements of 5 minutes acquisitions, the proposed MSMRI-PnP leads competitive performance even by exclusively using measurements of either 1 minute or 2 minute acquisitions.}
  \label{fig:exp}
\end{figure*}

\subsection{Using MSMRI-Net as an Image Prior}
CNN priors in the traditional PnP algorithm are designed for additive white gaussian noise (AWGN) removal.
In this section, we broaden the current denoiser-centric view by using a pre-trained MSMRI-Net as an image prior in PnP.
We called the resulting algorithm MSMRI-PnP and summarize it in Algorithm \ref{algo}. MSMRI-PnP can exploit the imaging prior specified for undersampled-artifacts, but can also systematically enforce data consistency between the solution and the raw measurements.
The pipeline of the MSMRI-PnP is also shown in Fig.\ \ref{fig:inference}.
We can use the formulation of PnP in~\eqref{equ:pnp} to express MSMRI-PnP:
\begin{equation}
  \xbm^k \leftarrow \textit{MSMRI-Net}(\xbm^{k-1}-\gamma \nabla g(\xbm^{k-1}))\ ,
\end{equation}
where we adapt the least-square penalty in~\eqref{equ:l2} and set initial vector $\xbm^{(0)}$ as zero-filled image of input undersampled measurement. We equip MSMRI-PnP with Nesterov's acceleration~\cite{nesterov2013introductory} via the sequence $\{q_k\}$ in Algorithm \ref{algo} for better convergence.  In particular, when $q_k=1$ for all $k$, the algorithms reverts to the usual method without acceleration.

\section{Experiment}
\subsection{Setup}
The proposed MSMRI-PnP framework were qualitatively and quantitatively evaluated on real in-vivo liver datasets.
All acquisition processes were performed on a 3T PET/MRI scanner (Biograph mMR; Siemens Healthcare, Erlangen, Germany). 
The data was collected using the CAPTURE method, a recently proposed T1-weighted stack-of-stars 3D spoiled gradient-echo sequence with fat suppression that has consistently acquired projections for respiratory motion detection~\cite{Eldeniz2018a}.
Upon the approval of our Institutional Review Board, multichannel liver data from 10 healthy volunteers and 6 cancer patients were used in this paper. 
We drop out the first 10 spokes during reconstruction in order to make sure the acquired signal reaches a steady state. 
Our free-breathing MRI data was then subsequently binned into 10 respiratory phases, and thus each phase view was reconstructed with 199 spokes accordingly. 
The coil sensitivity maps were estimated from the central radial k-space spokes of each slice and were assumed to be known during experiments. 
Apodization was applied by using a Hamming window that covers this range in order to avoid Gibbs ringing. 
We then implemented an inverse MCNUFFT on those individual coil element data and obtained 4D images of each subjects. 
Each image have $320\times 320\times 96\times 10$ dimensions with $320$ as both the width and the height, $96$ as the number of slice in the Sagittal direction and $10$ as the respiratory phases. 
We used the first 8 healthy subjects for our training and 1 for validation.
  
\subsection{Comparison}
Inspired by~\cite{Liu2019}, we compared the MSMRI-PnP with several widely used image reconstruction modalities, including CS and UNet3D. The CS implementation did 4D space-phase regularization using two complementary regularizers: 
(a) total variation (TV) that imposes piecewise smoothness across different respiratory phases and 
(b) total generalized variation (TGV) regularization that imposes higher-order piecewise smoothness in the spatial domain. 
Both regularizers are weighted by the regularization parameters that were optimized for the performance.
In particular, we considered CS reconstruction of maximum 5 minutes acquisition as high quality reference, termed as Reference.
UNet3D corresponds to conventional deep learning modality by mapping corrupted input to clean target directly.
Since maximum 5 minutes acquisition in our dataset are not long enough to obtain fully-sampled groundtruth, we exploited the Reference as the training target of the UNet3D.
Implementation of network structure and training strategy in the UNet3D are identical with the proposed MSMRI-Net to achieve a fair comparison.
We used peak signal-to-noise ratio (PSNR) in with respect to Reference to quantitatively estimate image reconstruction quality.

\subsection{Results}
Fig. \ref{fig:exp} shows visual reconstruction results of both 1 minutes and 2 minutes acquisitions.
Among those results, MCNUFFT reconstructions presents strong imaging noises and streaking artifacts while the other reconstruction both yield significant improvements.
Given image prior of spatial sparsity, CS reconstruction shows considerable reduction of the streaking artifact.
Despite that, it fails to gets rid of most image artifact, especially those imaging white noise.
The direct output of pre-trained MSMRI-Net can mitigate most imaging white noise, but it still suffers from few streaking artifact. 
By comparison with them, the proposed MSMRI-PnP stands out by leveraging the pre-trained CNN within physics-based iterative reconstruction and achieves competitive performance with UNet3D in terms of artifact removal, sharpness and contrast. 
More importantly, UNet3D requires measurements of 5 minutes acquisitions and an iterative CS modality to obtain clean target before training procedure, while the proposed MSMRI-PnP reconstructions of Fig. \ref{fig:exp} exclusively reply on measurement of either 1 minutes or 2 minutes acquisitions.
The value of PSNR labeled on top-right of each image also confirm these viewpoint.

\section{Conclusion}
We proposed a PnP based method for CS-MRI that systematically enforces data consistency and corroborate the potential of CNN priors trained from undersampled artifact-corrupted images. The CNN prior relies exclusively on downsampled variants of zero-filled images and reduces dependency of fully-sampled groundtruth.
We validated our modality on in-vivo acquisitions of undersampled 4D MRI data and observed that the proposed reconstruction of 1 or 2 minute acquisitions gained competitive performance compared with those CNN-based method required measurements of longer acquisition.

\section*{Acknowledgement}
Research reported in this publication was supported by the Washington University Institute of Clinical and Translational Sciences grant UL1TR002345 from the National Center for Advancing Translational Sciences (NCATS) of the National Institutes of Health (NIH). The content is solely the responsibility of the authors and does not necessarily represent the official view of the NIH.

\bibliographystyle{IEEEbib}

\end{document}